ARTICLE

Open Access

# 0.75 Gbit/s high-speed classical key distribution with mode-shift keying chaos synchronization of Fabry–Perot lasers

Hua Gao[1,2], Anbang Wang[1,2 ✉], Longsheng Wang[1,2], Zhiwei Jia[1,2], Yuanyuan Guo[1,2], Zhensen Gao[3,4], Lianshan Yan[5], Yuwen Qin[3,4] and Yuncai Wang[3,4]

**Abstract**
High-speed physical key distribution is diligently pursued for secure communication. In this paper, we propose and experimentally demonstrate a scheme of high-speed key distribution using mode-shift keying chaos synchronization between two multi-longitudinal-mode Fabry–Perot lasers commonly driven by a super-luminescent diode. Legitimate users dynamically select one of the longitudinal modes according to private control codes to achieve mode-shift keying chaos synchronization. The two remote chaotic light waveforms are quantized to generate two raw random bit streams, and then those bits corresponding to chaos synchronization are sifted as shared keys by comparing the control codes. In this method, the transition time, i.e., the chaos synchronization recovery time is determined by the rising time of the control codes rather than the laser transition response time, so the key distribution rate is improved greatly. Our experiment achieved a 0.75-Gbit/s key distribution rate with a bit error rate of $3.8 \times 10^{-3}$ over 160-km fiber transmission with dispersion compensation. The entropy rate of the laser chaos is evaluated as 16 Gbit/s, which determines the ultimate final key rate together with the key generation ratio. It is therefore believed that the method pays a way for Gbit/s physical key distribution.

## Introduction

Optical fiber communication has become the most important way for high-speed data transmission in modern communication networks, and then high-speed secure key distribution is urgently required to protect the transmitted data from eavesdropping. The algorithm-based key distribution has a risk of exhaustive attack due to the algorithm determinacy. Quantum key distribution promising unconditional security based on quantum no-cloning principle still has challenges: key rate is limited by the single-photon detector and distribution channel is hardly compatible with optical fiber communication links[1]. Therefore, classical physical methods have been inspired for high-speed secure key distribution.

The reported classical physical key distribution methods are mainly based on physical unclonable function[2], fiber laser[3–5], fiber channel noise[6–8], and electrical or optical chaos[9–17]. In the physical unclonable function method, two users extract random bits as private keys from the optical speckle pattern of each volumetric scattering material and form a public key dictionary[2]. Although the physical randomness of optical scattering supports a great number of random bits, the storage depth of public dictionary results in a finite length of the key sequence, which may lead to reuse of keys and risk of plaintext attack. For the fiber-laser method, legitimate users commonly modulate the laser wavelength or free spectral range with their independent random codes and then sift shared keys from the codes according to a certain laser state. The key rate determined by the state switching rate

Correspondence: Anbang Wang (wanganbang@tyut.edu.cn)
[1]Key Laboratory of Advanced Transducers and Intelligent Control System, Ministry of Education and Shanxi Province, Taiyuan, China
[2]College of Physics and Optoelectronics, Taiyuan University of Technology, Taiyuan, China
Full list of author information is available at the end of the article is inversely proportional to the transmission distance and thus lower than kbit/s[3]. In the method of fiber channel noise, legitimate users utilize optical interference[5,6] or mode mixing in multimode fiber[8] to generate noise signals with high correlation because of the channel reciprocity, and then extract shared keys from the correlated noise signals. The key distribution rate is limited to a few Mbit/s by the noise bandwidth[6], and the distance is usually about 25 km limited by the channel reciprocity.

Optical chaos in semiconductor lasers has been proposed for key distribution due to physical randomness which has been used to generate high-speed physical random bits[18–21], and due to chaos synchronization phenomenon[21–24]. Kanter et al. proposed using synchronous optical chaos of mutually coupled lasers to mask and exchange keys[9], and Porte et al. demonstrated a back-to-back experiment achieving a key rate of 11 Mbit/s[10]. The coupling leads to the transmission of chaotic carriers on public channel and thus could affect the security[25]. Yoshimura et al. proposed another key distribution scheme using common-signal-induced chaos synchronization, in which two uncoupled optical-feedback lasers are injected by a common light[11]. The security relies on that the information of entropy source can hardly be fully known by Eve because it is hard to get a laser with parameters well-matched with the user lasers. Yoshimura et al. further introduced random shift-keying modulation on the feedback phase as an additional layer of security[11]. A 182-kbit/s key distribution rate with 120-km fiber transmission was experimentally demonstrated by photonic-integrated optical-feedback lasers[13]. Subsequently, vertical-cavity surface-emitting laser under polarization-shift-keying optical injection[14] and analog–digital hybrid optical chaos source with phase-shift keying[15] were presented for key distribution. However, the final key rate of the laser-chaos-based distribution system with chaos-shift keying modulation is seriously restricted by the chaos synchronization recovery time in the magnitude of tens of nanoseconds which is limited by the laser relaxation oscillation frequency. It is noted that electrical chaos[16] and chaotic optoelectronic oscillator[17] have been demonstrated recently for key distribution in back-to-back synchronization but without keying-modulation on synchronization for an additional layer of security. For electrical chaos, its bandwidth limits the key rate. For optoelectronic oscillator, a short synchronization recovery time of 0.5 ns was numerically predicted[17], which indicates a high key rate but remains to be verified experimentally.

Here, we propose a novel key distribution scheme based on mode-shift keying chaos synchronization of multi-longitudinal-mode lasers. In this scheme, the chaos synchronization is achieved by the common injection of a super-luminescent diode (SLD) into Fabry–Perot (FP) lasers[22]. The keying chaos synchronization is realized by selecting laser mode rather than modulating the chaos state, avoiding the limitations of laser transition time on the chaos synchronization recovery time, and resultantly improving the rate of key distribution. The experimental demonstration shows that this method can increase the rate of key distribution based on chaos synchronization to 0.75 Gbit/s over a 160-km distance.

## Results
### Principle and system configuration

The schematic diagram of the proposed physical key distribution based on the mode-shift keying chaos synchronization of multi-longitudinal-mode lasers is illustrated in Fig. 1a. Two FP lasers with matched inner parameters are authorized to the legitimate users Alice and Bob. The two lasers are optically injected by a common random drive source and thus the modes with the same wavelength achieve chaos synchronization by adjusting optical injection parameters[22]. Then, dynamic optical filtering controlled by random binary codes $C_A$ ($C_B$) is applied to the laser $FP_A$ ($FP_B$), and the mode at wavelength $\lambda_0$ or $\lambda_1$ is filtered when $C_A$ ($C_B$) is a bit "0" or "1". As a result, Alice and Bob obtain their private chaotic waveforms with mode-shift keying, respectively. Only in the time slots when $C_A = C_B$, the filtered modes ($\lambda_0$ or $\lambda_1$ labeled red in Fig. 1a) are the same and the chaotic waveforms are synchronized. Subsequently, the two mode-shift keying chaotic signals are sampled and quantized to generate raw random bits $X_A$ and $X_B$, independently and privately. The random bits extracted during the time slots of chaos synchronization are identical in principle. Finally, Alice and Bob exchange and compare the control codes $C_{A,B}$ to sift identical bits as shared keys from $X_A$ and $X_B$, respectively. Note that more longitudinal modes can be used for the mode-shift keying, although only two modes are sketched in Fig. 1a.

The experimental system configuration is shown in Fig. 1b. An SLD is employed as the drive source. After filtering and amplification, the drive light is divided by a 3-dB optical coupler into two beams which are transmitted and then injected into two remote FP lasers. The transmission link between the two lasers consists of standard single-mode fibers and dispersion compensation fibers. At each user side, a variable optical attenuator and a polarization controller are used to adjust the optical power and the polarization state of the injected drive light, respectively. Here, two longitudinal modes are used for experimental demonstration. Dynamic optical filtering is implemented by the setup shown in the dashed box. A wavelength division multiplexer filters two modes of FP laser and splits them into two fiber paths, each with an electro-optic modulator (EOM). Binary control codes $C_A$ and the corresponding inverse codes (logical Not of $C_A$) generated from a random number generator are used to separately on/off modulate the two EOMs to achieve dynamic optical filtering. After optical coupling, the mode-shift keying chaos is obtained. An analog-to-digital converter in a real-time oscilloscope is used to sample the chaotic signal at a certain sampling rate to generate raw random bits by dual-threshold quantization[12]. By exchanging and comparing the control codes, the legitimate users sift the final keys from raw random bit sequences. Note that the drive source is not required to be strictly placed in the middle of the fiber link. The position offset of the drive source leads to a delay time which can be compensated by using a delay line in the user end closer to the drive source. Putting the drive source at one user end is feasible but the distance of key distribution could be shortened.

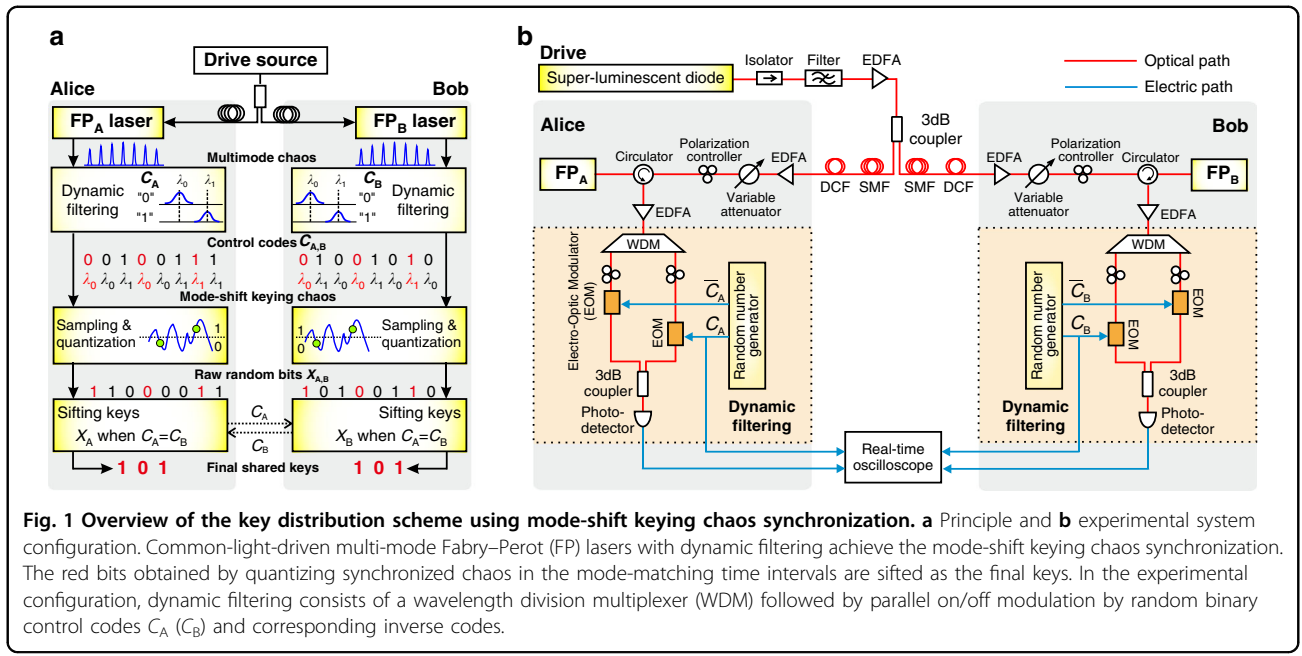

**Fig. 1 Overview of the key distribution scheme using mode-shift keying chaos synchronization. a** Principle and **b** experimental system configuration. Common-light-driven multi-mode Fabry–Perot (FP) lasers with dynamic filtering achieve the mode-shift keying chaos synchronization. The red bits obtained by quantizing synchronized chaos in the mode-matching time intervals are sifted as the final keys. In the experimental configuration, dynamic filtering consists of a wavelength division multiplexer (WDM) followed by parallel on/off modulation by random binary control codes $C_A$ ($C_B$) and corresponding inverse codes.

The scheme can realize the physical-layer security based on the following reasons. First, the outputs of the FP lasers cannot be intercepted in the fiber link because they are not transmitted and are low correlated to the drive signal. Second, it is hard for an attacker to obtain a third FP laser well-matched in inner parameters with the legitimate lasers because of the fabrication error[26]. With the above two conditions, eavesdropping information is less than the shared information of the legitimate users, while the drive signal is broadcasted, and thus security based on bounded observability[27] can be achieved, i.e., any attacker cannot get enough information of the entropy signals to obtain perfect duplicate keys. Third, random and private mode-shift keying further increases the difficulty of cracking. An attacker not only requires an FP laser but also needs to record the wideband chaotic signal of each wavelength of her own laser in order to guess which modes are employed by users. Simultaneous storage of massive high-speed data of several tens of modes brings out a challenge, which could lead to security based on bounded storage[27]. Therefore, the physical-layer security can be ensured.

## Chaos synchronization of FP lasers commonly driven by SLD

We first show the chaos synchronization properties in a back-to-back configuration. To achieve chaos synchronization, the center wavelengths, the relaxation oscillation frequencies, and the injection powers of the two FP lasers should be matched with each other. In our experiments, the lasers, $FP_A$ and $FP_B$, were biased at 1.14 and 1.13 times of threshold to ensure the same relaxation oscillation frequency of 2.64 GHz. Also, the lasers were injected by an SLD light with a spectral width of 5 nm and an injection power of 400 μW.

Figure 2a–d shows the single-longitudinal-mode chaos synchronization between two modes with the same wavelength $\lambda_0 = 1546.408$ nm. As shown in Fig. 2a and b, the modes of the two lasers have almost identical optical spectra and chaotic temporal waveforms. The scatter plot of the two chaotic waveforms in Fig. 2c concentrates around a straight line, meaning chaos synchronization. Quantitatively, the cross-correlation value, i.e. synchronization coefficient, is calculated as 0.9726 with a data length of 5000 ns. Figure 2d plots the histogram of short-term cross-correlation with a data length of 1 ns, and shows that the probability of the short-term cross-correlation values higher than 0.90 is 99.994%. This short-term robustness means that one can extract highly correlated random bits from the two chaotic lasers with a very high probability at a sampling rate of 1 Gbit/s at least, which is

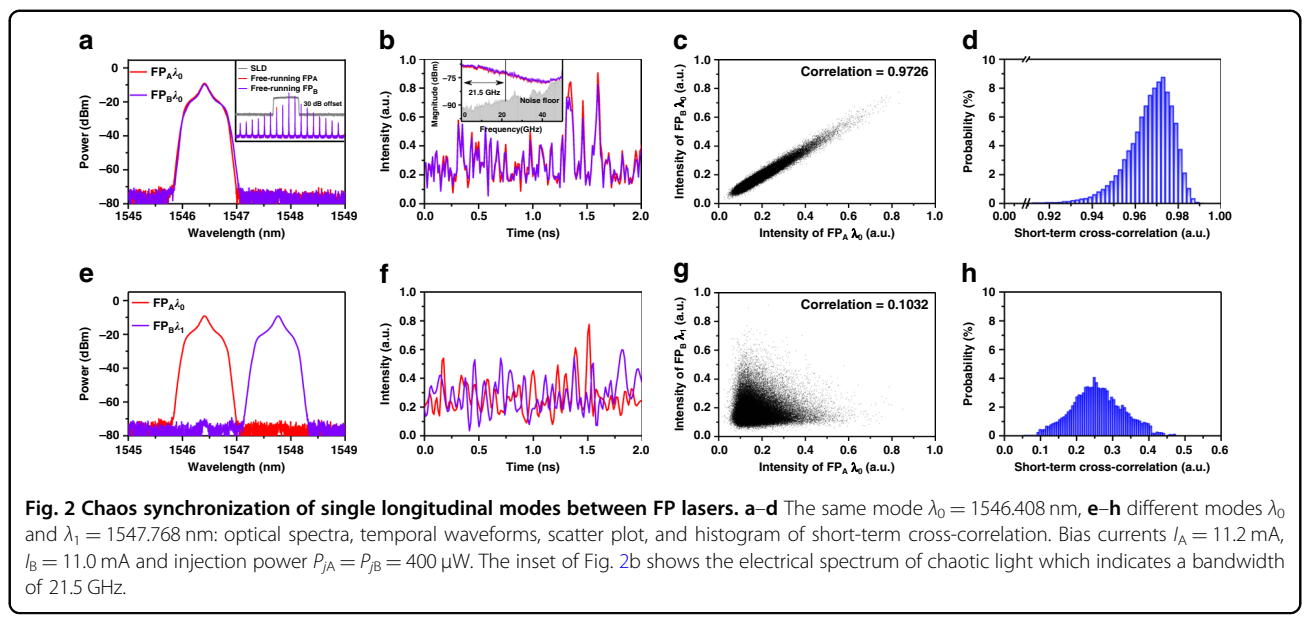

**Fig. 2 Chaos synchronization of single longitudinal modes between FP lasers. a–d** The same mode $\lambda_0 = 1546.408$ nm, **e–h** different modes $\lambda_0$ and $\lambda_1 = 1547.768$ nm: optical spectra, temporal waveforms, scatter plot, and histogram of short-term cross-correlation. Bias currents $I_A = 11.2$ mA, $I_B = 11.0$ mA and injection power $P_{jA} = P_{jB} = 400$ μW. The inset of Fig. 2b shows the electrical spectrum of chaotic light which indicates a bandwidth of 21.5 GHz.

sufficient for our experiments. The robustness mainly benefits from that the FP lasers have no external feedback cavity which is susceptible to environmental disturbance[12]. Long-term robustness or stability will be discussed in the next section. As plotted in the inset of Fig. 2b, the single-mode chaotic light has a flat and wide electrical spectrum. Evaluated as the span between the DC and the frequency where 80% energy is contained[28], the chaos bandwidth is ~21.5 GHz. The entropy rate of the chaotic light is estimated as 16 Gbit/s by the maximum generation rate of random bits with verified randomness using the single-bit quantization method (see Section S1 and Fig. S1 in Supplementary Information).

Figure 2e–h demonstrates a low cross-correlation value or non-synchronization between mode $\lambda_0$ of laser $FP_A$ and mode $\lambda_1 = 1547.768$ nm of laser $FP_B$. As shown in Fig. 2f and g, the chaotic temporal waveforms are obviously different, and the scatter plot does not exhibit a linear relationship. The cross-correlation value is only 0.1032. As shown in Fig. 2h, the probability of the short-term cross-correlation value lower than 0.35 is about 90%. The low correlation is mainly thanks to that the drive signals injected into the two modes are independent spontaneous emission noise signals which suppress the cross-gain modulation. Besides, the drive light and the FP lasers with the same filtering linewidth of 0.83 nm also have a low correlation measured as 0.38 (see Section S2 and Fig. S2 in Supplementary Information).

Note that in the inset of Fig. 2a, the drive light covers in spectral-domain four longitudinal modes of the free-running lasers. The synchronization coefficient of the four modes between the two lasers with injection was measured as 0.9678. This result indicates the feasibility of key distribution using multiple modes of lasers. Because more drive light noise is included, this synchronization coefficient is slightly lower than that of the single-mode.

Figure 3a–c depicts the effects of mismatch in external parameters including wavelength, injection power, and bias current on the cross-correlation of the same mode $\lambda_0$ between the two lasers, by adjusting one of the parameters of laser $FP_B$ while fixing the others. As plotted in Fig. 3a, the effects of wavelength mismatch exhibit symmetry and the correlation value is larger than 0.90 when the wavelength mismatch is within ±0.014 nm. As shown in Fig. 3b and c, when the injection power mismatch ranges from −20% to 25%, or when the current mismatch ranges from −5.4% to 7.1%, the correlation value keeps beyond 0.90. These tolerable mismatches show that the chaos synchronization is less sensitive to the external operation and indicate the robustness of synchronization. It is worth noticing that high sensitivity of chaos synchronization to mismatch in laser inner parameters is required to ensure security, which will be discussed later by numerical simulation.

### High-speed physical key distribution over 160-km fiber link

To demonstrate remote key distribution, two 80-km fiber links were separately configured between the driver and the two lasers, each of which consists of a 66-km standard single-mode fiber and a 14-km dispersion compensation fiber. The total distance was 160 km. After the fiber transmission, a synchronization coefficient of 0.9363 was achieved between modes with the same wavelength. Note that the lasers used in our experiment are monolithic lasers without external feedback cavity, and thus stable chaos synchronization can be achieved (see supplementary video of experimental chaos synchronization). We experimentally measured the cross-correlation between the two lasers within 90 min at an interval of 30 s, and a standard deviation of about 0.0009 was obtained (see Section S3 and Fig. S3 in Supplementary Information). It is noted that stable chaos synchronization can be achieved in commercial optical fiber networks due to that environmental perturbations such as vibrations and thermal expansion can be suppressed by using optical fiber cables. For example, a field experiment of chaos synchronization of semiconductor lasers with optical feedback had been successfully achieved in a 120-km commercial fiber link[29].

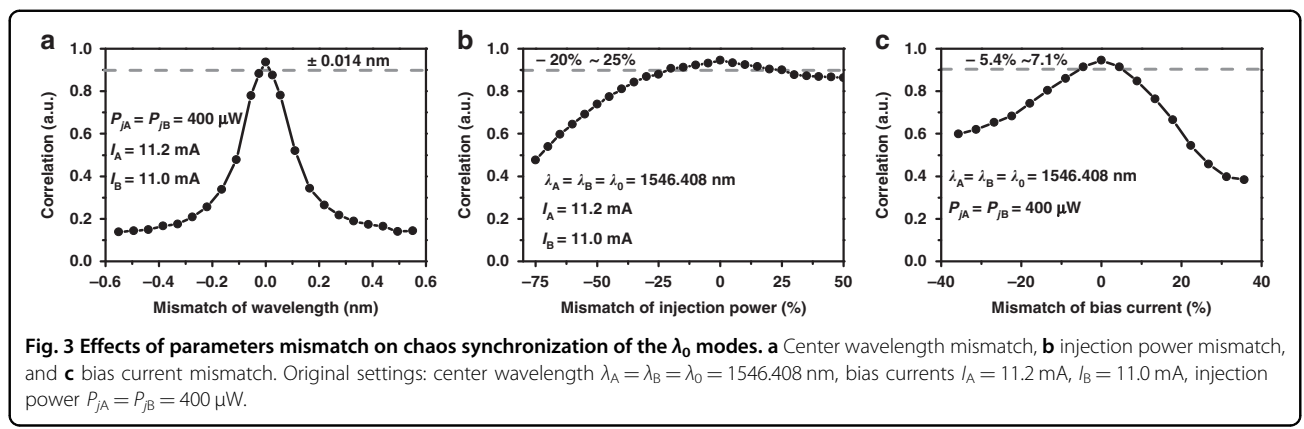

**Fig. 3 Effects of parameters mismatch on chaos synchronization of the $\lambda_0$ modes. a** Center wavelength mismatch, **b** injection power mismatch, and **c** bias current mismatch. Original settings: center wavelength $\lambda_A = \lambda_B = \lambda_0 = 1546.408$ nm, bias currents $I_A = 11.2$ mA, $I_B = 11.0$ mA, injection power $P_{jA} = P_{jB} = 400$ μW.

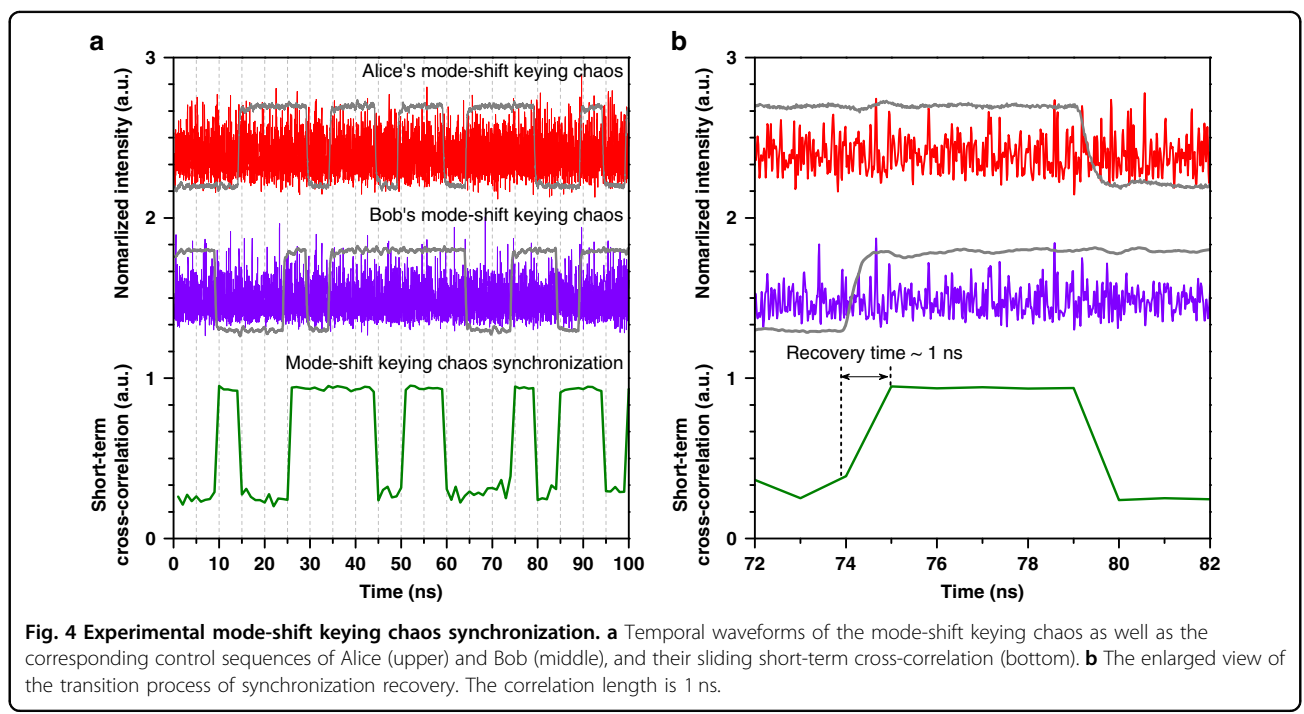

**Fig. 4 Experimental mode-shift keying chaos synchronization. a** Temporal waveforms of the mode-shift keying chaos as well as the corresponding control sequences of Alice (upper) and Bob (middle), and their sliding short-term cross-correlation (bottom). **b** The enlarged view of the transition process of synchronization recovery. The correlation length is 1 ns.

Then, we used two independent physical random bit generators with a rate of 200 Mbit/s based on Boolean chaos[30] to implement mode-shift keying for the two FP lasers, respectively. Figure 4 demonstrates typical experimental results of the mode-shift keying chaos synchronization. The temporal waveforms of the control codes and the corresponding mode-shift keying chaos of Alice and Bob are shown in the upper and the middle parts of Fig. 4a, respectively. The resultant mode-shift keying chaos synchronization is plotted at the bottom of Fig. 4a by the sliding short-term cross-correlation with a correlation length of 1 ns. Clearly, the synchronization coefficient reaches around 0.93 when the control codes are the same, but decreases to about 0.25 when the codes are different. In Fig. 4b, an enlarged view of the mode-shift keying chaos synchronization shows that the transition time from non-synchronization to synchronization is ~1 ns. This transition time is determined by the rising time of the electrical codes from the low level to the high level which is measured as 1 ns. It is greatly shorter than that of optical-feedback semiconductor lasers with phase-shift keying, which was reported as about 68 ns[13]. The shortened transition time is beneficial to improving the rate of key distribution.

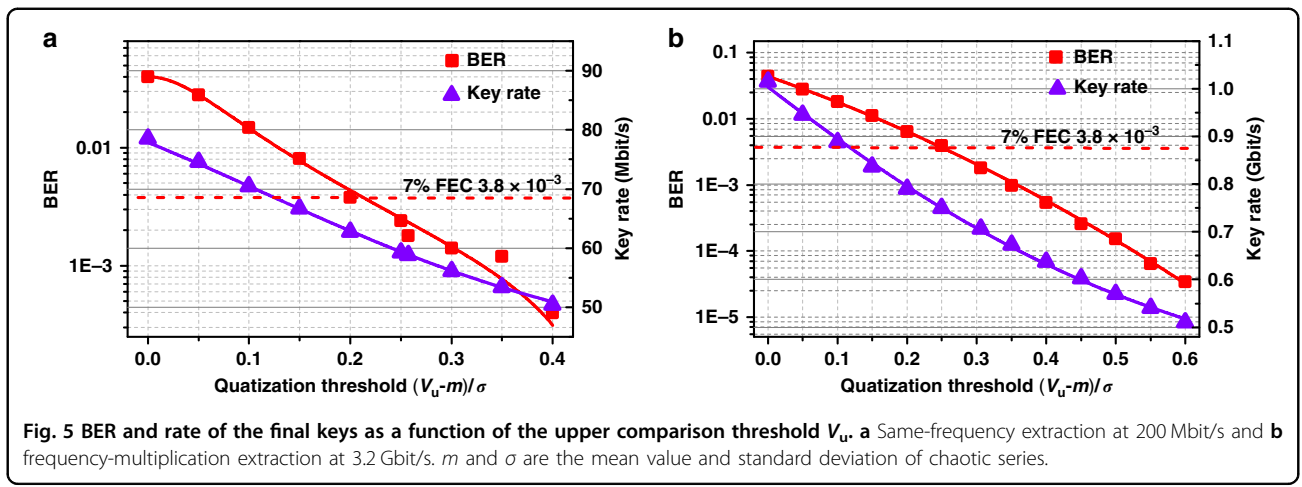

**Fig. 5 BER and rate of the final keys as a function of the upper comparison threshold $V_u$. a** Same-frequency extraction at 200 Mbit/s and **b** frequency-multiplication extraction at 3.2 Gbit/s. $m$ and $\sigma$ are the mean value and standard deviation of chaotic series.

Next, we used the two mode-shift keying chaotic waveforms to extract raw random bit sequences with dual-threshold quantization[12] and then sifted the final keys. Two voltages $V_u$ and $V_l$ were set as the upper and lower comparison threshold values, which can be normalized by $(V_{u,l} - m)/\sigma$, where $m$ and $\sigma$ are the mean and the standard deviation of the chaotic series. The sampled points were converted into bits "1" if their voltages were larger than $V_u$, or converted into bits "0" if smaller than $V_l$. The other sampled points distributed between $V_u$ and $V_l$ were discarded. The ratio of the number of raw bits to that of all sampled points is defined as the retention ratio of dual-threshold quantization. Among the raw bits, only the bits generated during time slots of chaos synchronization can be sifted out as the final keys. Thus, the key generation ratio is calculated as the ratio of the number of final keys to the number of sampled points and is also equal to the product of the retention ratio of quantization and the probability that the two shift-keying states are identical. The probability is 0.4982 for the used random generators in our experiment, which is close to the ideal value of 0.5. It is worth noticing that the stronger the system robustness is, the higher the retention ratio and the key rate become.

Figure 5a shows the bit error rate (BER) and the rate of the shared keys generated under an extraction rate of 200 Mbit/s equal to that of keying modulation. This case is named same-frequency extraction, by which only one point per keying-modulation period is sampled and quantized. Clearly, as the comparison threshold increases, both the BER and the key rate decrease, because more sampling points susceptible to noise are discarded. As normalized $V_u$ exceeds 0.204 (the corresponding $V_l$ is 0.445), the BER becomes smaller than the BER threshold of $3.8 \times 10^{-3}$ for hard-decision forward-error correction (HD-FEC) with 7% overhead, meaning that the error keys can be corrected by the FEC processing technique[31]. Therefore, a final key rate of 62.78 Mbit/s was achieved. The key rate is more than 300 times higher than that of the key distribution with phase-shift keying chaos synchronization[13] thanks to that the transition time is greatly shortened. The final key generation ratio is 0.3139, and the retention ratio of dual-threshold quantization is calculated as 0.6301, which is slightly higher than that of integrated optical-feedback semiconductor lasers[13] because the lasers without feedback have stronger robustness of chaos synchronization. According to the entropy rate of 16 Gbit/s and the key generation ratio of 0.3139, one can expect a 5 Gbit/s key distribution rate.

In order to increase the key rate, we sampled multiple points during each keying-modulation period to extract raw random bits. In this way, the sampling rate is equal to a multiple of the modulation frequency, which is called frequency-multiplication extraction. In this case, the sampling points in the 1-ns transition time window also should be discarded. By considering that a byte consists of 8 binary bits, the number of keys extracted from synchronized chaos during one modulation period should be <8 for security, because under this condition Eve can hardly crack a byte when a single modulation state is guessed. Figure 5b plots the experimental results of frequency-multiplication extraction with a sampling rate of 3.2 Gbit/s. It is noted that the key rate reaches 0.7503 Gbit/s with the BER of the FEC threshold under $V_u = 0.253$ and $V_l = 0.493$. The retention ratio of quantization is 0.4708 leading to about 7 bits per modulation period. Therefore, the security condition is satisfied. We further employed the NIST's statistical test suite[32] for

Table 1 Results of NIST test of key sequences generated from two FP lasers.

| Statistical test | FP$_A$ | | | FP$_B$ | | |
|---|---|---|---|---|---|---|
| | P-value | Proportion | Result | P-value | Proportion | Result |
| Frequency | 0.522100 | 0.9890 | Success | 0.461612 | 0.9860 | Success |
| Block frequency | 0.422638 | 0.9920 | Success | 0.585209 | 0.9900 | Success |
| Cumulative sums | 0.350485 | 0.9890 | Success | 0.471146 | 0.9870 | Success |
| Runs | 0.997147 | 0.9920 | Success | 0.721777 | 0.9890 | Success |
| Longest run | 0.994944 | 0.9870 | Success | 0.745908 | 0.9930 | Success |
| Rank | 0.878618 | 0.9880 | Success | 0.919131 | 0.9900 | Success |
| FFT | 0.420827 | 0.9840 | Success | 0.568739 | 0.9840 | Success |
| Non-overlapping template | 0.012043 | 0.9920 | Success | 0.006906 | 0.9860 | Success |
| Overlapping template | 0.937919 | 0.9860 | Success | 0.090388 | 0.9840 | Success |
| Universal | 0.138069 | 0.9890 | Success | 0.435430 | 0.9930 | Success |
| Approximate entropy | 0.814724 | 0.9940 | Success | 0.339271 | 0.9920 | Success |
| Random excursions | 0.048229 | 0.9891 | Success | 0.040363 | 0.9888 | Success |
| Random excursions variant | 0.008120 | 0.9922 | Success | 0.149903 | 0.9919 | Success |
| Serial | 0.137282 | 0.9910 | Success | 0.188601 | 0.9930 | Success |
| Linear complexity | 0.166260 | 0.9900 | Success | 0.140453 | 0.9890 | Success |

For tests that produce multiple P-values and proportions, the worst case is shown. Using 1000 samples of 1 Mbit data and significant level $\alpha = 0.01$, for "Success", the P-value should be larger than 0.0001 and the proportion should be in the range of $0.99 \pm 0.0094392$.

random number generators to test the generated keys. As shown in Table 1, all the 15 statistical tests were successfully passed, which verifies the randomness of the generated keys.

### Analysis of security

Chaos synchronization with parameter-matched lasers is the fundamental layer of security. The chaos synchronization should be highly sensitive to mismatch in laser inner parameters so that an attacker Eve can hardly get a third laser to achieve synchronization with the users' lasers. To examine this, we simulated the sensitivity of synchronization to the mismatch in parameters, including linewidth enhancement factor, active region length, interface reflection coefficient, linear gain coefficient, carrier density at transparency, and carrier capture time (see Section S4 in Supplementary Information). The maximum tolerable mismatch is about ±1% (Supplementary Fig. S4), requiring that two semiconductor lasers should be selected from the same fabrication wafer in order to achieve chaos synchronization[26]. In addition, this minor tolerable mismatch leads to a parameter space larger than $10^{16}$ (Section S4 in Supplementary Information). Thus, it is very hard for Eve to achieve the third laser with inner parameters matched with the legitimate lasers, under a reasonable assumption that Eve cannot access the same fabrication wafer which users have.

Therefore, the information per bit known by Eve from her laser with parameter mismatching as well as from the drive light about the shared bits of Alice and Bob is far smaller than 1, and thus secure keys can be generated[11].

The keying-modulation of synchronization provides an additional physical layer of security. In the method using the feedback-phase keying of DFB lasers with external optical feedback[11–13], the laser switches between different chaotic states. Thus, the information that Eve could know is further reduced, because more parameter-mismatched lasers are used to eavesdrop on different chaotic states. By comparison, in our method, the mode-shift keying modulation does not change the chaotic state and then cannot reduce Eve's information cracked by using lasers. It adds physical-layer security by boosting the difficulty in guessing which modes of lasers are used to generate keys by users.

### Discussion

In summary, we propose a novel physical key distribution based on mode-shift keying chaos synchronization and experimentally demonstrate a rate of 0.75 Gbit/s through 160-km fiber transmission with dispersion compensation. It is understood that chaotic waveforms in every transition period of keying modulation cannot be used to generate keys, and therefore shortening the transition time can increase the key generation rate. In this proposed method, the transition time or synchronization recovery time is greatly shortened, which is determined by the rising time of keying modulation. Utilizing a fast electrical pattern generator can further improve the key rate. In addition, the current result is a single-wavelength key rate because the keys were extracted from single-mode chaotic light. We notice that the multi-longitudinal-mode chaotic light of the two FP lasers can also be synchronized with wavelength matching. Thus, the key distribution rate can be still further improved through wavelength division multiplexing by using multiple modes for parallel generation of keys. It is believed that the proposed method has the capability of high-speed distribution with a key rate beyond 1 Gbit/s.

The current transmission distance can meet the requirement of metropolitan area networks. A longer transmission distance should be further researched for backbone networks. The long-distance transmission techniques in optical fiber communications such as Raman amplification could be used to increase the distance of key distribution. As discussed above, this method has the advantage of high speed and is compatible with the traditional fiber transmission systems. Therefore, it provides an alternative for high-speed secure communication, as a complementary technique for quantum key distribution.

## Materials and methods
### Chaos synchronization of FP lasers

The drive source SLD (Thorlabs SLD1005S) was biased at 400.0 mA and had a power of 13.56 mW with a 3-dB spectral width of 47.420 nm and a center wavelength of 1559.880 nm. The output light of SLD was filtered by a filter (Yenista XTM-50) with a linewidth of 5 nm and then amplified by an EDFA (Keopsys CEFA-C-HG) to 19.46 mW. Note that the filter of SLD can be removed because it does not affect the synchronization of single-mode light. The FP lasers (Junte GTLD-5FPBU10FA14) had similar parameters: threshold current $I_{thA} = 9.89$ mA and $I_{thB} = 9.73$ mA, slope efficiency of 0.248, and 0.265 mW/mA, mode interval of 1.36 nm. Their center wavelengths were matched by adjusting temperature controllers (ILX Lightwave LDT-5412) with a precision of 0.1 °C. Their bias currents were controlled as $1.14 I_{thA}$ and $1.13 I_{thB}$ by current sources (ILX Lightwave LDX-3412) to achieve the same relaxation oscillation frequency. The corresponding output powers were 325 and 336 μW.

The chaotic light was detected by a photodetector (Finisar XPDV2120R) with a bandwidth of 50 GHz and was measured by a real-time oscilloscope (Lecroy LABMASTER10ZI) with a bandwidth of 36 GHz and a sample rate of 80 GSa/s. A radio-frequency spectrum analyzer (Rohde&Schwarz FSW50) and an optical spectrum analyzer (Apex AP2041-B) were used to measure electrical spectra and optical spectra of the chaotic light.

### Mode-shift keying modulation

In experiments, the linewidth of each FP laser mode was broadened to about 0.082 nm due to optical injection. The linewidths of wavelength division multiplexers were 0.7 nm which was larger than that of the laser, which enables investigation on effects of wavelength mismatch on chaos synchronization. Two physical random bit generators based on Boolean chaos[29] were used to generate random non-return-to-zero binary codes for keying modulation. The random bit generation rate was 200 Mbit/s with a rising time of 1 ns and a high-level voltage of 2 V. The modulators (iXblue Photonics, MX-LN-40) have a bandwidth of 28 GHz and a half-wave voltage of 5.8 V.

### Random bit extraction

The 8-bit analog-to-digital converter in the real-time oscilloscope was utilized to quantize chaotic signals for random bit extraction. First, a chaotic signal was discretized and recorded by the oscilloscope with an 80 GSa/s sampling rate, which satisfies Nyquist's theorem for the chaos bandwidth of about 21.5 GHz. Then, for the random bit extraction with a rate of $f$ Gbit/s, a subsequence was reconstructed by picking the sampling points with a period equal to $80/f$ and then quantized by the dual-threshold method.

### Acknowledgements
This work was supported by the National Key R&D Program of China (2019YFB1803500), the National Natural Science Foundation of China (61822509, 62035009, 61731014, 61671316, 61805170), the Shanxi Talent Program (201805D211027), the Shanxi "1331 Project" Key Innovative Team, the Program for Top Young and Middle-aged Innovative Talents of Shanxi, and the Program for Guangdong Introducing Innovative and Entrepreneurial Teams.

### Author details
[1]Key Laboratory of Advanced Transducers and Intelligent Control System, Ministry of Education and Shanxi Province, Taiyuan, China. [2]College of Physics and Optoelectronics, Taiyuan University of Technology, Taiyuan, China. [3]School of Information Engineering, Guangdong University of Technology, Guangzhou, China. [4]Guangdong Provincial Key Laboratory of Photonics Information Technology, Guangzhou, China. [5]Center for Information Photonics and Communications, School of Information Science and Technology, Southwest Jiaotong University, Chengdu, China

### Author contributions
A.B.W. and H.G. conceived the experiment. A.B.W. guided and H.G. implemented the experimental work and data analysis. H.G. made the first draft writing. A.B.W., L.S.W., H.G., Z.W.J., Y.Y.G., and Z.S.G. contributed to the final version of the manuscript. A.B.W., L.S.Y., Y.W.Q., and Y.C.W. supervised the project.

### Data availability
The data that support the plots within this paper and other findings in this study are available from the corresponding author upon reasonable request.

### References

1. Liao, S. K. et al. Satellite-to-ground quantum key distribution. *Nature* **549**, 43–47 (2017).
2. Horstmeyer, R. et al. Physical key-protected one-time pad. *Sci. Rep.* **3**, 3543 (2013).
3. Scheuer, J. & Yariv, A. Giant fiber lasers: a new paradigm for secure key distribution. *Phys. Rev. Lett.* **97**, 140502 (2006).
4. El-Taher, A. et al. Secure key distribution over a 500 km long link using a Raman ultra-long fiber laser. *Laser Photonics Rev.* **8**, 436–442 (2014).
5. Tonello, A. et al. Secret key exchange in ultralong lasers by radiofrequency spectrum coding. *Light: Sci. Appl.* **4**, e276 (2015).
6. Kravtsov, K. et al. Physical layer secret key generation for fiber-optical networks. *Opt. Express* **21**, 23756–23771 (2013).
7. Hajomer, A. A. E. et al. Key distribution based on phase fluctuation between polarization modes in optical channel. *IEEE Photonics Technol. Lett.* **30**, 704–707 (2018).
8. Bromberg, Y. et al. Remote key establishment by random mode mixing in multimode fibers and optical reciprocity. *Opt. Eng.* **58**, 016105 (2019).
9. Kanter, I. et al. Synchronization of random bit generators based on coupled chaotic lasers and application to cryptography. *Opt. Express* **18**, 18292–18302 (2010).
10. Porte, X. et al. Bidirectional private key exchange using delay-coupled semiconductor lasers. *Opt. Lett.* **41**, 2871–2874 (2016).
11. Yoshimura, K. et al. Secure key distribution using correlated randomness in lasers driven by common random light. *Phys. Rev. Lett.* **108**, 070602 (2012).
12. Koizumi, H. et al. Information-theoretic secure key distribution based on common random-signal induced synchronization in unidirectionally-coupled cascades of semiconductor lasers. *Opt. Express* **21**, 17869–17893 (2013).
13. Sasaki, T. et al. Common-signal-induced synchronization in photonic integrated circuits and its application to secure key distribution. *Opt. Express* **25**, 26029–26044 (2017).
14. Jiang, N. et al. Secure key distribution based on chaos synchronization of VCSELs subject to symmetric random-polarization optical injection. *Opt. Lett.* **42**, 1055–1058 (2017).
15. Zhao, Z. X. et al. Semiconductor-laser-based hybrid chaos source and its application in secure key distribution. *Opt. Lett.* **44**, 2605–2608 (2019).
16. Keuninckx, L. et al. Encryption key distribution via chaos synchronization. *Sci. Rep.* **7**, 43428 (2017).
17. Böhm, F. et al. Stable high-speed encryption key distribution via synchronization of chaotic optoelectronic oscillators. *Phys. Rev. Appl.* **13**, 064014 (2020).
18. Uchida, A. et al. Fast physical random bit generation with chaotic semiconductor lasers. *Nat. Photonics* **2**, 728–732 (2008).
19. Li, P. et al. Ultrafast fully photonic random bit generator. *J. Lightwave Technol.* **36**, 2531–2540 (2018).
20. Kanter, I. et al. An optical ultrafast random number bit generator. *Nat. Photonics* **4**, 58–61 (2010).
21. Sciamanna, M. & Shore, K. A. Physics and applications of laser diode chaos. *Nat. Photonics* **9**, 151–162 (2015).
22. Tomiyama, M. et al. Effect of bandwidth limitation of optical noise injection on common-signal-induced synchronization in multi-mode semiconductor lasers. *Opt. Express* **26**, 13521–13535 (2018).
23. Oliver, N., Jüngling, T. & Fischer, I. Consistency properties of a chaotic semiconductor laser driven by optical feedback. *Phys. Rev. Lett.* **114**, 123902 (2015).
24. Chembo, Y. K. et al. Optoelectronic oscillators with time-delay feedback. *Rev. Mod. Phys.* **91**, 035006 (2019).
25. Antonik, P. et al. Using a reservoir computer to learn chaotic attractors, with applications to chaos synchronization and cryptography. *Phys. Rev. E* **98**, 012215 (2018).
26. Argyris, A. et al. Chaos-on-a-chip secures data transmission in optical fiber links. *Opt. Express* **18**, 5188–5198 (2010).
27. Muramatsu, J. et al. Secret-key distribution based on bounded observability. *Proc. IEEE* **103**, 1762–1780 (2015).
28. Lin, F. Y., Chao, Y. K. & Wu, T. C. Effective bandwidths of broadband chaotic signals. *IEEE J. Quantum Electron.* **48**, 1010–1014 (2012).
29. Argyris, A. et al. Chaos-based communications at high bit rates using commercial fibre-optic links. *Nature* **438**, 343–346 (2005).
30. Zhang, R. et al. Boolean chaos. *Phys. Rev. E* **80**, 045202 (2009).
31. Argyris, A., Bueno, J. & Fischer, I. Photonic machine learning implementation for signal recovery in optical communications. *Sci. Rep.* **8**, 8487 (2018).
32. Bassham, L. et al. *A Statistical Test Suite for Random and Pseudorandom Number Generators for Cryptographic Applications* (NIST Special Publication, 2010).